\documentclass[10pt,english,pre,nofootinbib,reprint,showpacs,showkeys,preprintnumbers,floatfix]{revtex4-1}
\usepackage[latin9]{inputenc}
\usepackage[a4paper]{geometry}
\usepackage{babel}
\usepackage{units}
\usepackage{amsmath}
\usepackage{amssymb}
\usepackage{graphicx}
\usepackage{esint}
\usepackage[unicode=true,pdfusetitle,
 bookmarks=true,bookmarksnumbered=false,bookmarksopen=false,
 breaklinks=false,pdfborder={0 0 1},backref=false,colorlinks=false]
 {hyperref}

\makeatletter
 
 \@ifundefined{textcolor}{}
 {%
   \definecolor{BLACK}{gray}{0}
   \definecolor{WHITE}{gray}{1}
   \definecolor{RED}{rgb}{1,0,0}
   \definecolor{GREEN}{rgb}{0,1,0}
   \definecolor{BLUE}{rgb}{0,0,1}
   \definecolor{CYAN}{cmyk}{1,0,0,0}
   \definecolor{MAGENTA}{cmyk}{0,1,0,0}
   \definecolor{YELLOW}{cmyk}{0,0,1,0}
 }

\usepackage{aas_macros}
\usepackage{bbold}

\makeatother

\begin{document}

\title{The Galaxy in circular polarization: \\
all-sky radio prediction, detection strategy, \\
and the charge of the leptonic cosmic rays }

\author{Torsten A. Enßlin}
\altaffiliation{Max Planck Institute for Astrophysics, Karl-Schwarzschildstr.1, 85741 Garching, Germany;
Ludwig-Maximilians-Universität München, Geschwister-Scholl-Platz 1, 80539 Munich, Germany}

\selectlanguage{english}%

\author{Sebastian Hutschenreuter}
\altaffiliation{Max Planck Institute for Astrophysics, Karl-Schwarzschildstr.1, 85741 Garching, Germany;
Ludwig-Maximilians-Universität München, Geschwister-Scholl-Platz 1, 80539 Munich, Germany}

\selectlanguage{english}%

\author{}

\author{Valentina Vacca}
\altaffiliation{Osservatorio Astronomico Cagliari, Via della Scienza 5 - 09047 Selargius, Italy}

\selectlanguage{english}%

\author{Niels Oppermann}
\altaffiliation{Canadian Institute for Theoretical Astrophysics; 60 St. George Street; Toronto, ON M5S 3H8; Canada
Dunlap Institute for Astronomy and Astrophysics; 50 St. George Street; Toronto, ON M5S 3H4; Canada}

\selectlanguage{english}%
\begin{abstract}
The diffuse Galactic synchrotron emission should exhibit a low level
of diffuse circular polarization (CP) due to the circular motions
of the emitting relativistic electrons. This probes the Galactic magnetic
field in a similar way as the product of total Galactic synchrotron
intensity times Faraday depth. We use this to construct an all sky
prediction of the so far unexplored Galactic CP from existing measurements.
This map can be used to search for this CP signal in low frequency
radio data even prior to imaging. If detected as predicted, it would
confirm the expectation that relativistic electrons, and not positrons,
are responsible for the Galactic radio emission. Furthermore, the
strength of real to predicted circular polarization would provide
statistical information on magnetic structures along the line-of-sights. 
\end{abstract}
\maketitle

\section{Introduction}

\subsection{Circular polarisation emission}

The radio synchrotron emission of the Milky Way should be circularly
polarized due to the circular motions of relativistic electrons in
the Galactic magnetic field (GMF). Because of the relativistic beaming
effect of the electron's motion on the emitted radiation we see mostly
the electrons that spiral around fields oriented perpendicular to
the line-of-sight (LOS) and therefore predominantly linearly polarized
emission. The magnetic fields that point towards us could be a source
of circular polarization (CP), reflecting the circular motions of
the relativistic electrons visible in this geometry. However, the
aforementioned beaming effect diminishes any radiation parallel to
the magnetic field. The largest CP emission should therefore result
from magnetic fields with an inclination in between parallel and perpendicular
to the LOS. The field component parallel to the LOS, $B_{\|}$ , ensures
that a circular component of the electron gyration is visible to the
observer, and determines thereby the sign of Stokes-$V$. The field
component perpendicular to the LOS, $B_{\perp}$, enables the gyrating
electrons to send some beamed flux into the direction of the observer
and therefore largely determines the strength of the CP. 

So far only linear polarization has been detected and imaged in the
diffuse radio-synchtron emission of the Milky Way \citep{2006A&A...448..411W,2007ApJS..170..335P,2016ApJ...822...34P}.\footnote{CP from compact Galactic objects like Sagrittarius\ $\text{A}^{*}$,
GRS\ 1915, SS\ 433 has been detected \citep{1999ApJ...523L..29B,1999ApJ...526L..85S,2000ApJ...530L..29F,2002MNRAS.336...39F},
which seems to result from a different process as discussed here,
namely Faraday conversion operating in the much stronger magnetic
fields of these objects \citep{1977ApJ...214..522J,1977ApJ...215..236J,1982ApJ...263..595H,1984Ap&SS.100..227V,1984MNRAS.208..409K,1990MNRAS.242..158B,2002A&A...388.1106B,2002ApJ...573..485R,2003A&A...401..499E}.} CP should be much weaker and therefore harder to be detected and
charted. Nevertheless, Galactic CP emission should exist and therefore
should in principle be observable. Since the CP signal is weak, and
has to be discriminated from instrumental polarization leakage effects,
it would be very helpful to have a prediction not only on the magnitude
of this emission, but also on its detailed morphology on the sky.
This paper provides such a prediction.

\subsection{Predicting circular polarisation}

To predict the CP emission accurately knowledge of the GMF strength
and orientation is necessary throughout the Galactic volume, as well
on the number density and the energy spectrum of the relativistic
electron population. Currently we are lacking this information, despite
substantial efforts to model the GMF \citep{2008A&A...477..573S,2009A&A...495..697W,2009JCAP...07..021J,2010MNRAS.401.1013J,2010RAA....10.1287S,2011A&A...526A.145F,2011MNRAS.416.1152J,2012ApJ...757...14J,2012ApJ...761L..11J,2016A&A...596A.103P},
the Galactic thermal electrons \citep{2002astro.ph..7156C,2016A&A...590A..59G},
and relativistic electrons \citep{2001AdSpR..27..717S,2004ApJ...613..962S,2011CoPhC.182.1156V,2012A&A...542A..93O,2015arXiv150705958O,2016APh....81...21M,2017JCAP...02..015E}
from various observables. The observables informing us about the perpendicular
GMF component (times the relativistic electron density) are the linear
polarization and total emission of the synchrotron emission. The parallel
GMF component imprints onto the Galactic Faraday rotation measures
of extra-galactic sources, however modulated by the thermal electron
density. Instead of using these observables to construct a 3D GMF
model from which CP can be predicted \citep{2016PhRvD..94b3501K},
here we exploit that a certain combination of these observables should
be linearly correlated with the CP signal. Exploiting this correlation
without the detour of building a simplified 3D GMF model should permit
to predict more small-scale structures of the Galactic CP signal than
by usage of a coarse 3D GMF model. The CP sky prediction by \citep{2016PhRvD..94b3501K}
is based on such a 3D model and exhibits small scale structures. The
latter are, however, due to a random magnetic field added to the 3D
model, and therefore will not represent the real small-scale structures
of the CP sky.

The small scale-structures of our predicted CP signal will be more
realistic. However, they only represent a statistical guess for the
real CP signal. The pieces of information that are put together, the
Faraday signal as a tracer of $B_{\|}$ and the total synchrotron
intensity as a tracer of $B_{\perp}$, might report about different
locations along the LOS, whereas the combination of both components
at the locations of CP emission would be needed. The former signal
results predominately from locations of high thermal and the latter
of high relativistic electron density, and these do not need to coincide
spatially. 

Fortunately, the GMF exhibits some spatial correlation as the observables
are correlated as a function of sky direction and this correlation
should also hold in the LOS direction. Therefore, information on field
components resulting from slightly different locations might still
provide a good guess at a position. Some of the structures imprinted
onto the observables are caused by structures in the underlying thermal
or \textendash{} to a lesser degree \textendash{} relativistic electron
population and might, however, be misleading and lead to spurious
structures in a CP sky predicted this way. 

Anyhow, a CP prediction constructed directly form such observables
will be mostly model independent and therefore ideal for template-based
CP detection efforts. It will have small angular scale structures
that should also permit the usage of interferometer data that usually
lack large angular scale sensitivity. While using this CP template
map, it should just be kept in mind that it resembles an educated
guess for the galactic CP morphology, and is certainly not accurate
in all details. It should, however, be a very helpful template for
the extraction of a detectable signal out of the probably noisy CP
data and help to verify the detection of the Galactic CP signal by
discriminating this from instrumental systematics that plague the
measurement of weak polarization signals.

\subsection{Testing the charge of the emitters}

The rotational sense of the CP flux should typically have the opposite
sign of that of the Faraday rotation if both are measured with the
same convention and if the relativistic and thermal particles involved
have the same charge sign, e.g. are both electrons. The reason is
that the CP sense should directly reflect the gyro-motion of the relativistic
particles emitting the radio emission. Faraday rotation is caused
by the different phase speeds of left and right polarised electromagnetic
waves in a magnetized plasma. The waves that co-rotates with the lightest
thermal charge carriers \textendash{} usually the electrons \textendash{}
can interact most strongly with them and gets the largest delay. Consequently,
a linear wave that can be regarded as a superposition of left and
right circular waves gets rotated in the sense of the faster wave,
and therefore counter rotates with respect to the gyro-motion of the
light charge carriers. Thus, if the involved thermal and relativistic
particles have the same charge sign, CP and Faraday rotation produced
in the same magnetic field counter rotate. This opens the possibility
to test for the existence of regions with positrons dominating the
radio synchrotron emission.

\subsection{Structure of the paper}

The structure of the paper is the following. Sect.\ \ref{sec:Circular-polarization-Sky}
presents the theoretical derivation of the CP prediction map construction.
Sect.\ \ref{sec:Prediction} provides the predicted CP sky and discusses
its remaining model uncertainties. Sect.\ \ref{sec:Detection-strategy}
investigates the detectability of the predicted signal and Sect.\ \ref{sec:Conclusions}
concludes.

\section{Circular polarization Sky\label{sec:Circular-polarization-Sky}}

\subsection{Observables}

The CP intensity as characterized by Stokes $V$ for a given LOS is
approximately given by
\begin{equation}
V=\alpha_{V}\int\!\!dl\,n_{\mathrm{rel}}B_{||}\,B_{\perp}^{3/2},
\end{equation}
 where 

\begin{equation}
\alpha_{V}=-\frac{0.342\cdot\mathrm{e}^{9/2}}{\pi\,\sqrt{2\pi}\,\nu^{3/2}\,m_{\mathrm{e}}c^{7/2}(\gamma_{\mathrm{min}}^{-2}-\gamma_{\mathrm{max}}^{-2})}\label{eq:alpha_v}
\end{equation}
is a constant, which depends only on natural constants, model parameters
and the CP-observational frequency $\nu$ \citep{2016ApJ...822...34P}.
The symbols $\mathrm{e},\,m_{e}$ and $c$ denote the elementary charge,
the electron mass, and the speed of light, respectively. The relativistic
electrons with density $n_{\mathrm{rel}}$ are assumed here to have
the same power law-like spectrum with cut offs $\gamma_{\mathrm{min}}$
and $\gamma_{\mathrm{max}}$ and spectral index of $p_{\mathrm{e}}=-3$
everywhere. $B_{||}$ is the LOS-parallel, and $B_{\perp}$ the LOS-perpendicular
magnetic field component and we wrote $\int\!\!dl=\int_{\mathrm{LOS}}dl$
for the LOS integration. Also by writing $B_{\perp}^{3/2}$ we assumed
implicitly an electron power law index of $p_{e}=-3$, which is not
too far from the one observed. However, this simplification could
be dropped if needed by replacing $B_{\perp}^{3/2}$ with $B_{\perp}^{\nicefrac{-p_{e}}{2}}$
everywhere, for the price of more contrived calculations. As we only
strive for a rough estimate of the CP signal, we will continue with
the simpler $B_{\perp}^{3/2}$ scaling. 

The building blocks of the CP signal $n_{\mathrm{rel}}$, $B_{||}$,
and $B_{\perp}$ appear in nearly the same combination in the Galactic
total synchrotron intensity $I$ and the Faraday depth $\phi,$
\begin{align}
I & =\alpha_{I}\int\!\!dl\,n_{\mathrm{rel}}B_{\perp}^{2},\label{eq:I}\\
\phi & =\alpha_{\phi}\int\!\!dl\,n_{\mathrm{th}}B_{||}.
\end{align}
Here, 
\begin{equation}
\alpha_{I}=\frac{\mathrm{e}^{4}}{6\pi m_{\mathrm{e}}^{2}c^{3}\nu\,(\gamma_{\mathrm{min}}^{-2}-\gamma_{\mathrm{max}}^{-2})}\label{eq:alpha_i}
\end{equation}
\citep{2016ApJ...822...34P} and 
\begin{equation}
\alpha_{\phi}=\frac{e^{3}}{2\pi m_{e}^{2}c^{4}}
\end{equation}
\citep{2012A&A...542A..93O} are other constants, which depend on
similar natural constants and model parameters as $\alpha_{V}$, and
$n_{\mathrm{th}}$ is the density of free thermal electrons. In particular,
the data combination
\begin{equation}
d=\phi\,I=\alpha_{\phi}\alpha_{I}\int\!\!dl\int\!\!dl'\,n_{\mathrm{th}}(l)\,n_{\mathrm{rel}}(l')\,B_{||}(l)\,B_{\perp}^{2}(l')
\end{equation}
contains the same magnetic field components as $V=\alpha_{V}\int dl\,n_{\mathrm{rel}}(l)\,B_{||}(l)\,B_{\perp}^{3/2}(l)$,
although with slightly different spatial dependence and a slightly
different $B_{\perp}$ dependence. If the magnetic field would be
spatially constant along a LOS, $d$ and $V$ would be correlated
according to
\begin{equation}
\frac{V}{d}=\frac{\alpha_{V}}{\alpha_{\phi}\alpha_{I}}\frac{1}{\int\!\!dl\,n_{\mathrm{th}}\,B_{\bot}^{1/2}}\;,
\end{equation}
so that knowing $d$ would allow us to predict $V$ ap art from the
weak $B_{\bot}$ dependence, assuming we know the LOS integrated thermal
electron density from other measurements like pulsar dispersions.
In reality, $d$ and $V$ will not be perfectly correlated as there
are unknown magnetic structures on the LOS. The ratio
\begin{equation}
\frac{V}{d}=\frac{\alpha_{V}}{\alpha_{\phi}\alpha_{I}}\,\frac{\int dl\,n_{\mathrm{rel}}B_{||}\,B_{\perp}^{3/2}}{\left(\int\!\!dl\,n_{\mathrm{th}}B_{||}\right)\left(\int\!\!dl\,n_{\mathrm{rel}}B_{\perp}^{2}\right)}
\end{equation}
therefore encodes information on magnetic structures along the LOS,
in particular on the co-spatiality of Faraday rotating and synchrotron
emitting regions. This information would be interesting to obtain
in order to improve our GMF models. 

Before CP observations can be exploited for studying Galactic magnetism,
the CP signal has to be detected. For this, a rough model of the CP
sky would be extremely helpful, as it can be used to build optimal
detection templates to be applied to the noisy CP data. In the following,
we construct such a predictive CP-polarization all sky map for this
purpose. As $d$ is already an observable today, it can be used to
predict $V$ to some degree. 

$V$ and $d$ will in general be correlated. The production of CP
is inevitably associated with total intensity emission and the sign
of the produced $V$ is determined by the sign of $B_{||}$, which
always also imprints into the Faraday depth (for emission locations
with thermal electrons). This correlation might be weak, in case the
synchrotron emission and Faraday depth signals are mostly created
at distinct locations with mostly uncorrelated magnetic LOS component
$B_{||}$. If, on the other hand, synchrotron emissions and Faraday
rotation are mainly co-spatial, a strong correlation between $V$
and $d$ can be expected. The fact that the Galactic radio emission
exhibits strong signatures of Faraday depolarization \textbf{\citep{2006A&A...448..411W}}
supports the idea of an intermixed Faraday rotating and synchrotron
emitting medium, which promises a large cross-correlation of $d$
and $V$. Thus the prospects for predicting the CP sky signal to some
degree are good.

\subsection{Model}

All three observables under consideration here, $I$, $\phi,$ and
$V$, could be predicted for a given Galactic model in $n=(n_{\mathrm{th}},n_{\mathrm{rel}})$
and $\vec{B}=(B_{||},\vec{B}_{\perp})$, where we have chosen the
LOS direction to be always our first coordinate. Although we have
rough models for the 3D Galactic electron distributions $n$, the
full 3D GMF configuration is currently poorly known. The existing
GMF models \citep{2008A&A...477..573S,2009A&A...495..697W,2009JCAP...07..021J,2010MNRAS.401.1013J,2010RAA....10.1287S,2011A&A...526A.145F,2011MNRAS.416.1152J,2012ApJ...757...14J,2012ApJ...761L..11J,2016A&A...596A.103P}
largely exploit the available Faraday and synchrotron data and therefore
do not contain too much in addition to what these data-sets have to
offer. The additional information of these models is due to the usage
of parametric models of the GMF spiral structure, which are inspired
from the observations of other galaxies. Although this is certainly
helpful information, the price to be paid for it is a loss of small-scale
structure in the model prediction as the parametric models do not
capture all complexity of the data sets they are fitted to. These
small-scale structures are, however, extremely important for detecting
the Galactic CP signal, as many radio telescopes and in particular
radio interferometers are insensitive to large-scale angular structures.
Furthermore, a GMF model based prediction is only superior on large
scales if the included additional assumptions were correct. Although,
this might well be the case, to have a more model independent prediction
is certainly healthy. 

For these reasons, we will try to predict the CP sky from existing
$I$ and $\phi$ sky maps directly, using only a minimal set of absolutely
necessary model assumptions, which we describe now. The inclusion
of more information and assumptions is in principle possible and would
to lead more sophisticated $V$-map predictions as we are aiming for
here. 

As the fluctuations in our observables are mainly caused by magnetic
field structures and to a lesser degree by structures in the electron
densities $n=(n_{\mathrm{th}},\,n_{\mathrm{rel}})$, for which rough,
but sufficiently accurate models exist, we will assume $n$ to be
known along any given LOS. For $n_{\mathrm{th}}$ we adopt the large-scale
structure of the popular NE2001 model \citep{2002astro.ph..7156C}
and $n_{\mathrm{rel}}$ is modeled as a thick exponential discs, with
parameters as specified in detail in Sec. \ref{sec:Prediction}. Adapting
a simplistic model for the electron densities means that any structure
in the RM sky, which is a consequence of not modeled structures in
the thermal electron density, will be attributed to magnetic field
structures and imprints on the resulting CP sky. Thus, the predicted
CP sky will show some features not being present in the real CP sky.
Not modeled structures in the relativistic electron density will imprint
to both, the total intensity map and the CP map. Therefore, those
will imprint on the CP prediction despite the fact that the inference
model assigns them to magnetic sub-structures internally.

Although the detailed GMF is still a matter of research, reasonable
guesses for how the magnetic energy density scales typically with
Galactic locations as expressed through $n$ exist and will be adopted
here. This means, we assume that the GMF energy density is largely
a function of the electron density. We therefore need an expression
for 
\begin{equation}
\overline{B}^{2}(n)=\langle\vec{B}^{2}\rangle_{(\vec{B}|n)}
\end{equation}
with $\langle f(x,y)\rangle_{(x|y)}=\int\!\!dx\,\mathcal{P}(x|y)\,f(x,y)$
expressing the probabilistic expectation value of a function $f(x,y)$
(here $\vec{B}^{2}$) averaged over the conditional probability $\mathcal{P}(x|y)$
of an unknown variable $x$ (here $\vec{B}$) given a known variable
$y$ (here $n$ to characterize the different typical environments
in the Galaxy).

In this work, a simple parametrization of the form 
\begin{equation}
\overline{B}^{2}(n)=\frac{B_{0}^{2}}{n_{\mathrm{th}0}^{\beta_{\mathrm{th}}}n_{\mathrm{rel}0}^{\beta_{\mathrm{rel}}}}n_{\mathrm{th}}^{\beta_{\mathrm{th}}}n_{\mathrm{rel}}^{\beta_{\mathrm{rel}}}=B_{0}^{2}\,x_{\mathrm{th}}^{\beta_{\mathrm{th}}}x_{\mathrm{rel}}^{\beta_{\mathrm{rel}}}\label{eq:scaling-relation}
\end{equation}
will be used, with $x_{i}\equiv\nicefrac{n_{i}}{n_{i0}}$ and plausible
scaling indices of $\beta=(\beta_{\mathrm{th}},\beta_{\mathrm{rel}})\in[0,1]^{2}$
. To be definitive, we adopt $\beta_{\mathrm{th}}=0$ and $\beta_{\mathrm{rel}}=1$
to model our intuition that the observed thick synchrotron disk of
the Milky Way and other galaxies probably require magnetic fields
which have a thick disk as well as the relativistic electrons causing
this thick disk emission. This is in line with the expectation that
the relativistic fluid in galaxies, consisting of mainly of relativistic
protons, other ions, and electrons, drags magnetic fields with it
when it streams out of galactic disks.

In order to show to which degree our CP sky prediction depends on
this assumption we also show results for the complementary case $\beta=(1,0)$.
It will turn out that $\beta$ has only a marginal effect on our prediction,
indicating also that the 3D modeling of the electron distributions
is not the most essential input to our calculation. The exact normalization
of the scaling relation Eq.\ \ref{eq:scaling-relation} is given
by the parameters $B_{0}^{2},n_{\mathrm{th}0}^{\beta_{th}}$ and $n_{\mathrm{rel}0}^{\beta_{rel}}$.
In the explicit calculation later on we use $B_{0}\approx6\,\mu\mathrm{G}$
and $n_{\mathrm{th}0}\approx5\cdot10^{-2}\mathrm{cm^{-3}}$. The parameter
for the relativistic electron density $n_{\mathrm{rel}0}$ drops out
later on in the course of the calculation and is therefore left unspecified.
The reason for this is that it affects the observable $I$ in exactly
the same way as the predicted quantity $V$, and therefore becomes
irrelevant when conditioning our prediction on the observable $I$,
which contains the necessary information on $n_{\mathrm{rel}0}$ . 

We will exploit the correlation of $V$ with the quantity $d=\phi I$
to predict the former. These quantities depend on the magnetic field
structure along a LOS in different ways. Their cross-correlation depends
on the magnetic field correlation tensor

\begin{equation}
M_{ij}(\vec{x},\vec{y})=\langle B_{i}(\vec{x})\,B_{j}(\vec{y})\rangle_{(\vec{B})}
\end{equation}
as well as on higher correlations functions. A priori, we have no
reason to assume that within a roughly homogeneous Galactic environment
(as defined by roughly constant $n$) any direction or location to
be singled out. Thus, a statistical homogeneous, isotropic, and mirror-symmetric
correlation tensor should model our a priori knowledge about the field,
which then is of the form \citep{1999PhRvL..83.2957S}
\begin{eqnarray}
M_{ij}(\vec{x},\vec{y}) & = & M_{ij}(\vec{r})\\
 & = & M_{\text{N}}(r)\,\delta_{ij}+\left(M_{\text{L}}(r)-M_{\mathrm{\text{N}}}(r)\right)\,\hat{r}_{i}\hat{r_{j}},\nonumber 
\end{eqnarray}
with $M_{\mathrm{N}}(r)$ and $M_{\mathrm{L}}(r)$ normal and longitudinal
scalar correlation functions, which depend only on the magnitude $r$
of the distance vector $\vec{r}=\vec{x}-\vec{y}$ with normalized
components $\hat{r}_{i}=r_{i}/r$. These functions describe the correlation
of the field at one location with that at another location shifted
in a normal or longitudinal direction with respect to the local magnetic
field orientation. These correlation functions are connected due to
$\vec{\nabla}\cdot\vec{B}=0$ via
\begin{equation}
M_{\mathrm{\text{N}}}(r)=\frac{1}{2r}\frac{d}{dr}\left[r^{2}\,M_{\text{L}}(r)\right]
\end{equation}
and can be combined into the magnetic scalar correlation $w(r)=\langle\vec{B}(\vec{x})\cdot\vec{B}(\vec{x}+\vec{r})\rangle_{(\vec{B})}=2M_{\mathrm{\text{N}}}(r)+M_{\text{L}}(r)$
so that $\overline{B}^{2}=w(0)=2M_{\text{N}}(0)+M_{\text{L}}(0)$
\citep{1999PhRvL..83.2957S}. 

In our calculations, only correlations along of LOSs are needed, leading
to the restriction $\vec{r}=(r,0,0)$ if we identify the LOS direction
with the first coordinate axis. This implies a component-wise diagonal
correlation structure
\begin{eqnarray}
\left.M_{ij}(\vec{r})\right|_{\vec{r}=(r,0,0)} & = & \left[M_{\mathrm{\text{N}}}(r)+\left(M_{\text{L}}(r)-M_{\text{N}}(r)\right)\,\delta_{i1}\right]\,\delta_{ij}\nonumber \\
 & = & \begin{pmatrix}M_{\text{L}} & 0 & 0\\
0 & M_{\text{N}} & 0\\
0 & 0 & M_{\text{N}}
\end{pmatrix}_{ij}\!\!\!\!(r)\label{eq:M-diagonal}
\end{eqnarray}
and therefore no a priori expectation of any cross-correlation of
$B_{||}$ and $B_{\perp}$ along a given LOS. This simplifies the
calculation of higher order magnetic correlation functions. For such
we will use the Wick theorem, e.g. 
\[
\langle B_{i}B_{j}B_{k}B_{l}\rangle_{(\vec{B})}=M_{ij}M_{kl}+M_{ik}M_{jl}+M_{il}M_{jk},
\]
and therefore implicitly a Gaussian probability for the magnetic field
components. The real magnetic field statistics is most likely non-Gaussian,
leading to differences between our estimated higher order correlates
and the real ones. However, since we do not know how to model this
non-Gaussianity correctly as we do not know even the sign of its effect
on higher order correlations, and as we also like to keep the complexity
of our calculations moderate we accept this simplification. We expect
only a moderate and global multiplicative change of order unity on
our predicted CP sky if the nature of non-Gaussianity would be known
and taken into account in the prediction, as non-Gaussianity corrections
would roughly affect all LOSs more or less similarly. 

Furthermore, we assume the longitudinal and normal magnetic correlation
lengths (defined here differently to match our later needs)
\begin{eqnarray}
\lambda_{\text{L}} & = & \int\!\!dr\,M_{\mathrm{\text{L}}}(r)/M_{\text{L}}(0)\mbox{ and}\nonumber \\
\lambda_{\text{N}} & = & \int\!\!dr\,M_{\text{N}}^{2}(r)/M_{\mathrm{\text{N}}}^{2}(0)\label{eq:lambda}
\end{eqnarray}
to be much smaller than typical variations in the underlying electron
density profiles, so that e.g. the expected Farday dispersion can
be calculated via 
\begin{eqnarray}
\langle\phi^{2}\rangle_{(\vec{B}|n)} & = & \alpha_{\phi}^{2}\int_{0}^{\infty}\!\!\!\!\!\!\!dl\,\int_{0}^{\infty}\!\!\!\!\!\!\!dl'\,n_{\mathrm{th}}(l)\,n_{\mathrm{th}}(l')\langle B_{||}(l)\,B_{||}(l')\rangle_{(\vec{B}|n)}\nonumber \\
 & \approx & \alpha_{\phi}^{2}\int_{0}^{\infty}\!\!\!\!\!\!\!dl\,\int_{-\infty}^{\infty}\!\!\!\!\!\!\!dr\,n_{\mathrm{th}}(l)\,n_{\mathrm{th}}(l+r)\,M_{\text{L}}(r)\nonumber \\
 & \approx & \frac{1}{3}\alpha_{\phi}^{2}\lambda_{\text{L}}\int_{0}^{\infty}\!\!\!\!\!\!\!dl\,n_{\mathrm{th}}^{2}\,\overline{B}^{2}(n).\label{eq:phiphi}
\end{eqnarray}
We introduced the notation $f(l)=f(l\,\widehat{r}_{\mathrm{LOS}})$
for the value of the 3D field $f(\vec{r})$ along the LOS coordinate
$l$ in direction $\widehat{r}_{\mathrm{LOS}}$ . Here, and in the
following we will treat the individual LOSs separately. Furthermore,
we assumed that magnetic structures are smaller than the part of the
LOS that resides in the Galaxy as expressed in terms of the structure
of the adopted thermal electron model, so that a negligible error
is implied by extending the integration over the relative distances
$r=l'-l$ from minus to plus infinity or by using the same thermal
electron density for both locations, $l$ and $l+r$. Furthermore,
we used $M_{\text{L}}(0)=M_{\text{N}}(0)=\frac{1}{3}\overline{B}^{2}$,
which follows from isotropy and Eq.\ \ref{eq:M-diagonal}.

Finally, we assume the observed Faraday and total intensity skies
to be noiseless. This approximation will simplify the CP sky estimator
and make it independent of the normalization of the scaling relation
Eq.\ \ref{eq:scaling-relation} and the actual value of the correlation
length $\lambda_{\text{L}}$ as long this does not vary (strongly)
along a given LOS. The assumed correlations length $\lambda_{\text{N}}$
will have some small impact on our result, however, of sub-dominant
order and therefore it is also not necessary to specify it if only
a rough CP sky prediction is required.

\subsection{Estimator}

\begin{figure*}
\raggedright{}\includegraphics[bb=70bp 0bp 372bp 216bp,clip,width=0.5\textwidth]{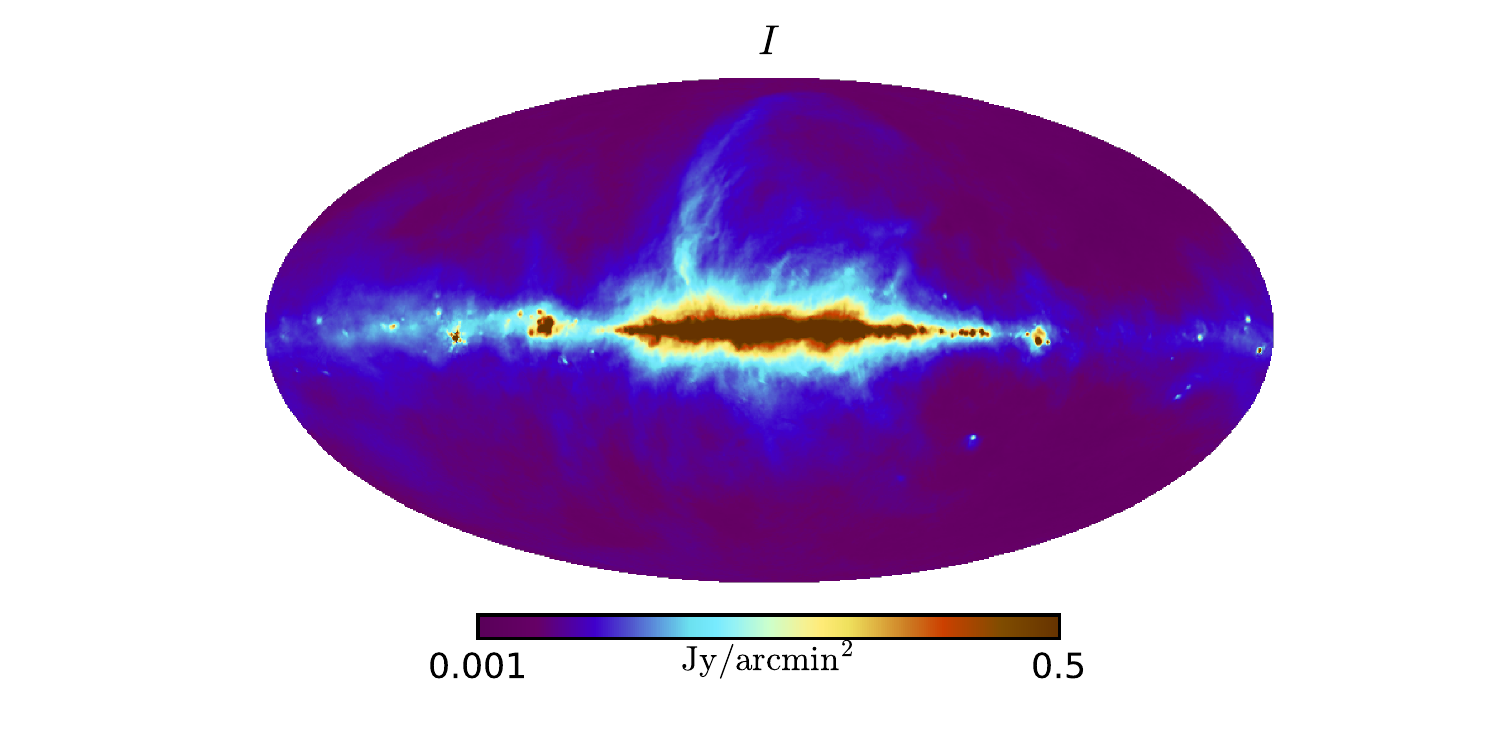}\includegraphics[bb=70bp 0bp 372bp 216bp,clip,width=0.5\textwidth]{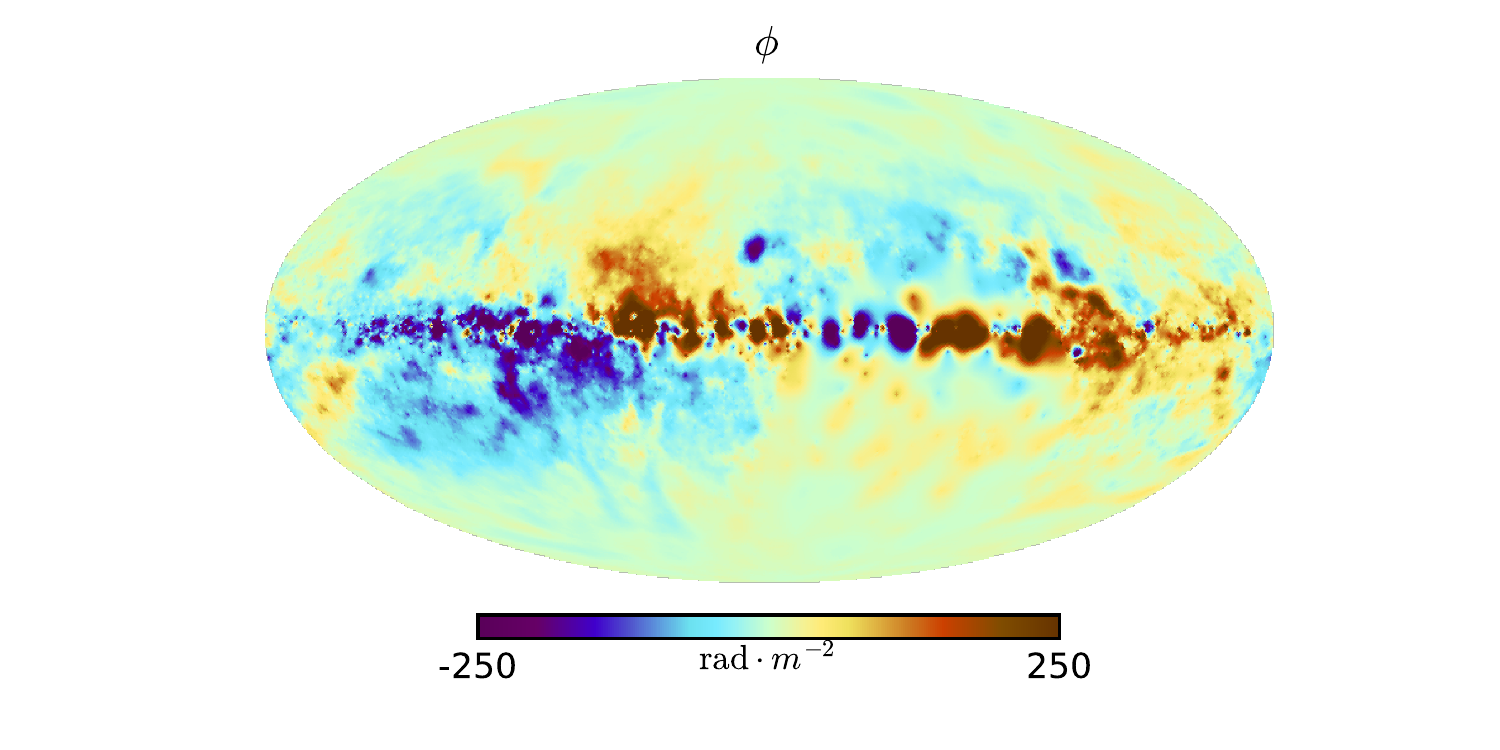}\caption{Left: Synchrotron intensity at 408 MHz as provided by \citep{2015MNRAS.451.4311R}.\label{fig:haslam}
Right: Faraday rotation map as constructed by \citep{2012A&A...542A..93O}.
Red indicates magnetic fields predominantly pointing towards the observer
and clockwise rotation of the received linear polarisation. This is
according to the IAU convention for measuring angles and is therefore
opposite to the mathematical convention.\label{fig:faraday} }
\end{figure*}
\begin{figure*}
\includegraphics[bb=70bp 0bp 372bp 216bp,clip,width=0.5\textwidth]{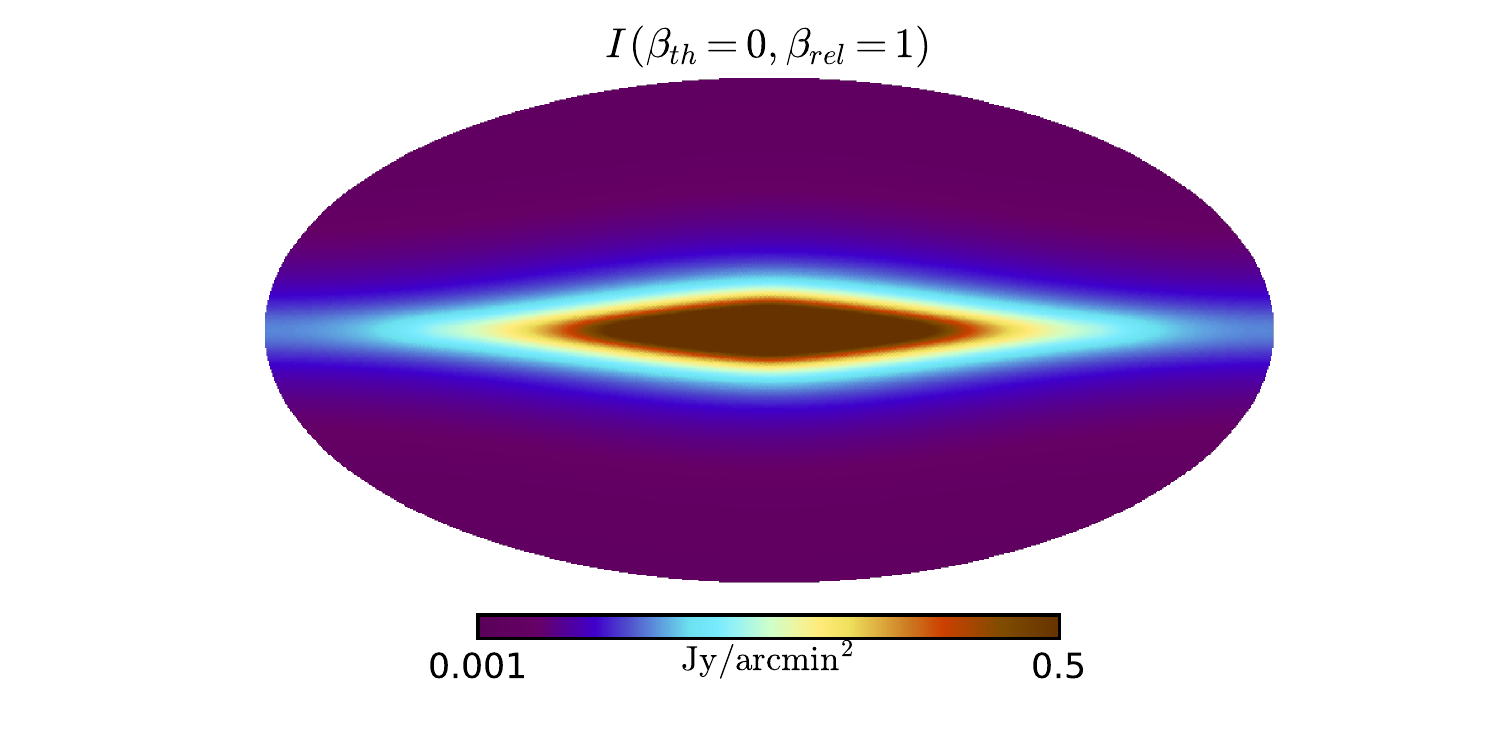}\includegraphics[bb=70bp 0bp 372bp 216bp,clip,width=0.5\textwidth]{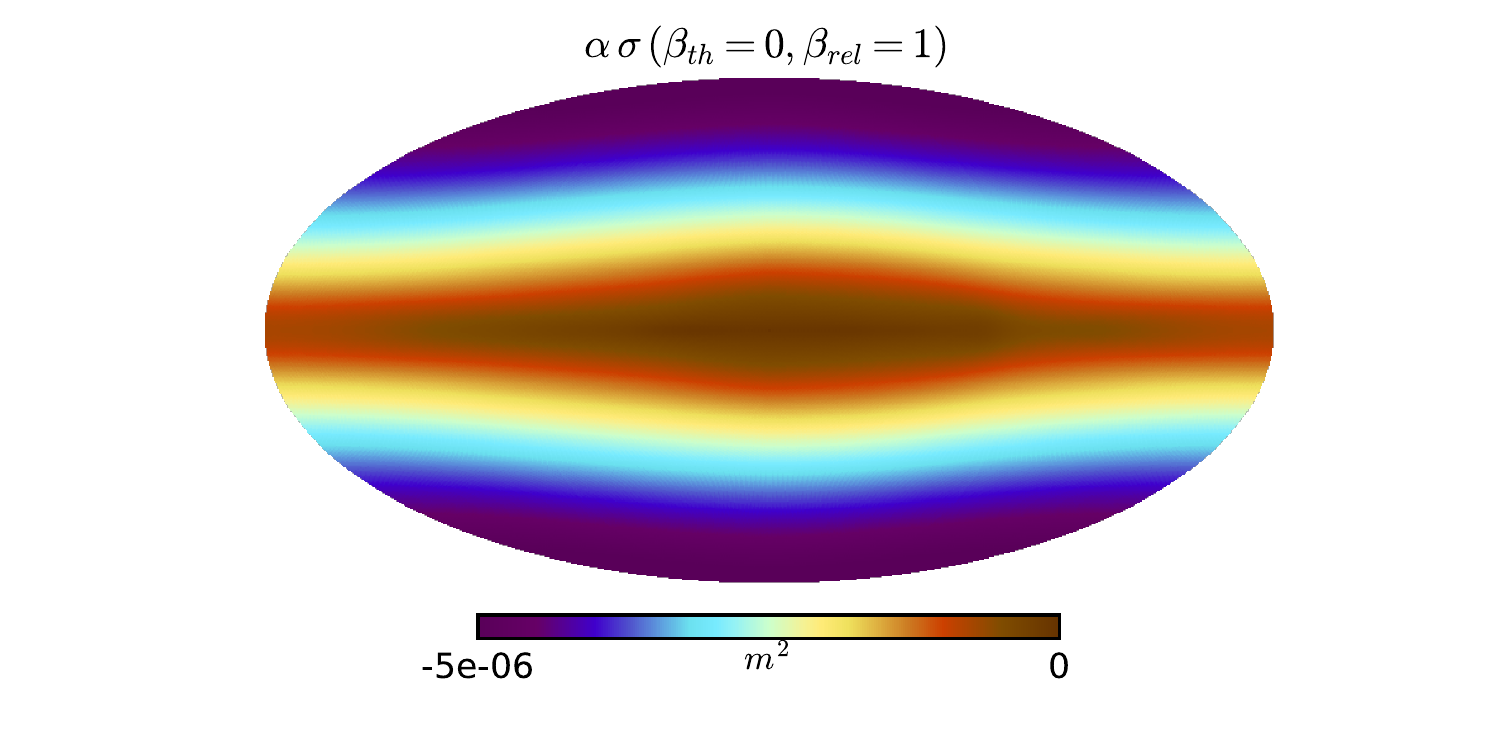}\caption{Left: Synchrotron emission intensity at 408 MHz of the simplistic
3D model. Right: Map of the resulting conversion factor $\alpha\sigma$,
which translates the Faraday rotation map $\phi$ into the fractional
CP map $V/I$ at 408 MHz. For both the relativistic electron profile
of Eq.\ \ref{eq:x_rel} and $\beta=(0,1)$ were assumed. \label{fig:i_predict_01}}
\end{figure*}
We want to exploit the correlation of $V$ with $d=\phi\,I$ to construct
an optimal linear estimator for $V$ given $d$. This is given by
\begin{equation}
\overline{V}=\langle V\,d\rangle_{(\vec{B}|n)}\langle d^{2}\rangle_{(\vec{B}|n)}^{-1}d\label{eq:WF}
\end{equation}
irrespectively the underlying statistics, since one can easily show
that the quadratic error expectation $\epsilon^{2}=\langle\left[V-\overline{V}(d)\right]^{2}\rangle_{(\vec{B}|n)}$
is always minimized for linear estimators of the form $\overline{V}(d)=v\,d$
for $v=\langle V\,d\rangle_{(\vec{B}|n)}\langle d^{2}\rangle_{(\vec{B}|n)}^{-1}$:
\begin{eqnarray}
\frac{d\epsilon^{2}}{dv} & = & -2\langle\left[V-v\,d\right]\,d\rangle_{(\vec{B}|n)}\nonumber \\
 & = & 2\,\left[v\langle d^{2}\rangle_{(\vec{B}|n)}-\langle V\,d\rangle_{(\vec{B}|n)}\right]=0.
\end{eqnarray}

All remaining analytical work is to calculate the correlates which
compose $v.$ The simpler one is 
\begin{eqnarray}
\langle d^{2}\rangle_{(\vec{B}|n)} & = & \langle\phi^{2}I^{2}\rangle_{(\vec{B}|n)}\nonumber \\
 & = & \alpha_{\phi}^{2}\alpha_{I}^{2}\int\!\!dl_{1}\ldots\int\!\!dl_{4}\,n_{\mathrm{th}1}\,n_{\mathrm{th}2}\,n_{\mathrm{rel}3}\,n_{\mathrm{rel}4}\times\nonumber \\
 &  & \langle B_{||1}B_{||2}B_{\perp3}^{2}B_{\perp4}^{2}\rangle_{(B|n)}\nonumber \\
 & = & \alpha_{\phi}^{2}\alpha_{I}^{2}\int\!\!dl_{1}\ldots\int\!\!dl_{4}\,n_{\mathrm{th}1}\,n_{\mathrm{th}2}\,n_{\mathrm{rel}3}\,n_{\mathrm{rel}4}\times\nonumber \\
 &  & M_{\text{L}12}\,\left[M_{\text{N}33}\,M_{\text{N}44}+2\,M_{\text{N}34}^{2}\right]\nonumber \\
 & \approx & \frac{1}{27}\,\lambda_{\text{L}}\alpha_{\phi}^{2}\alpha_{I}^{2}\,\left[\int\!\!dl\,n_{\mathrm{th}}^{2}\overline{B}^{2}\right]\times\nonumber \\
 &  & \left[\left(\int\!\!dl\,n_{\mathrm{rel}}\,\overline{B}^{2}\right)^{2}+2\lambda_{\text{N}}\int\!\!dl\,n_{\mathrm{rel}}^{2}\,\overline{B}^{4}\right].
\end{eqnarray}
Here, we used the abbreviations $n_{\mathrm{th}1}=n_{\mathrm{th}}(l_{1})$,
$B_{||2}=B_{||}(l_{2})$, $M_{\text{N}34}=M_{\text{N}}(l_{3}-l_{4})$,
and the like, exploited the diagonal structure of the magnetic correlations
along the LOS as expressed by Eq.\ \ref{eq:M-diagonal} while applying
the Wick theorem, and inserted the correlation lengths $\lambda_{\text{L}}$
and $\lambda_{\text{N}}$ as defined in Eq.\ \ref{eq:lambda} while
applying the short correlation length approximation as previously
used in Eq.\ \ref{eq:phiphi}.

The calculation of $\langle V\,d\rangle_{(\vec{B}|n)}$ is slightly
more complicated. To handle the $B_{\bot}^{3/2}$ dependence of $V$,
we Taylor expand it in terms of $B_{\bot}^{2}$ around $B_{\bot0}^{2}=\frac{2}{3}B_{0}^{2}$
via
\begin{flalign}
B_{\bot}^{\nicefrac{3}{2}} & =(B_{\bot}^{2})^{\nicefrac{3}{4}}=\underset{n=0}{\sum^{\infty}}\binom{\nicefrac{3}{4}}{n}\,B_{\bot0}^{2(\frac{3}{4}-n)}\left(B_{\bot}^{2}-B_{\bot0}^{2}\right)^{n}\nonumber \\
 & =\binom{\nicefrac{3}{4}}{0}\,B_{\bot0}^{\nicefrac{3}{2}}+\binom{\nicefrac{3}{4}}{1}\,B_{\bot0}^{-\frac{1}{2}}\left(B_{\bot}^{2}-B_{\bot0}^{2}\right)+\mathcal{O\left(\mathrm{\mathit{B}}_{\bot}^{\mathrm{4}}\right)}\nonumber \\
 & \approx\frac{1}{4}B_{\bot0}^{\nicefrac{3}{2}}+\frac{3}{4}\,B_{\bot0}^{-\nicefrac{1}{2}}\,B_{\bot}^{2}.
\end{flalign}

We choose to expand in $B_{\perp}^{2}$ rather than $B_{\perp}$,
as the linear terms would vanish anyway during the application of
the Wick theorem. 

We then find: 
\begin{eqnarray}
\langle V\,d\rangle_{(\vec{B}|n)} & = & \langle V\,\phi\,I\rangle_{(\vec{B}|n)}\nonumber \\
 & = & \alpha_{V}\alpha_{\phi}\alpha_{I}\int\!\!dl_{1}\ldots\int\!\!dl_{3}\,n_{\mathrm{th}1}\,n_{\mathrm{rel}2}\,n_{\mathrm{rel}3}\times\nonumber \\
 &  & \langle B_{||1}B_{||2}B_{\perp2}^{\frac{3}{2}}B_{\perp3}^{2}\rangle_{(B|n)}\nonumber \\
 & \approx & \alpha_{V}\alpha_{\phi}\alpha_{I}\int\!\!dl_{1}\ldots\int\!\!dl_{3}\,n_{\mathrm{th}1}\,n_{\mathrm{rel}2}\,n_{\mathrm{rel}3}\times\nonumber \\
 &  & \!\!\!\!\!\!\!\!\!\!\!\!\!\!\left\langle B_{||1}B_{||2}\,\left(\frac{1}{4}B_{\bot0}^{\nicefrac{3}{2}}+\frac{3}{4}\,B_{\bot0}^{-\nicefrac{1}{2}}\,B_{\bot2}^{2}\right)B_{\perp3}^{2}\right\rangle _{(B|n)}\nonumber \\
 & = & \alpha_{V}\alpha_{\phi}\alpha_{I}\int\!\!dl_{1}\ldots\int\!\!dl_{3}\,n_{\mathrm{th}1}\,n_{\mathrm{rel}2}\,n_{\mathrm{rel}3}\times\nonumber \\
 &  & \frac{B_{\bot0}^{-\nicefrac{1}{2}}}{4}\,M_{\text{L}12}\,\bigg(B_{\bot0}^{2}M_{\text{N}33}+\nonumber \\
 &  & \mbox{}\mbox{}+3\,\left[M_{\text{N}22}\,M_{\text{N}33}+2\,M_{\text{N}23}^{2}\right]\bigg)\nonumber \\
 & \approx & \frac{B_{\bot0}^{-\frac{1}{2}}}{36}\,\lambda_{\text{L}}\alpha_{V}\alpha_{\phi}\alpha_{I}\,\times\nonumber \\
 &  & \biggl[B_{\bot0}^{2}\left(\int\,\,dl\,n_{\mathrm{th}}n_{\mathrm{rel}}\,\overline{B}^{2}\right)\left(\int\,\,dl\,n_{\mathrm{rel}}\,\overline{B}^{2}\right)\nonumber \\
 &  & +\left(\int\!\!dl\,n_{\mathrm{th}}\,n_{\mathrm{rel}}\,\overline{B}^{4}\right)\left(\int\!\!dl\,n_{\mathrm{rel}}\,\overline{B}^{2}\right)+\nonumber \\
 &  & 2\lambda_{\text{N}}\int\!\!dl\,n_{\mathrm{th}}\,n_{\mathrm{rel}}^{2}\,\overline{B}^{6}\biggr].
\end{eqnarray}

Again we used $M_{\text{L}}(0)=M_{\text{N}}(0)=\frac{1}{3}\overline{B}^{2}$
and $\lambda_{\text{L}}$ and $\lambda_{\text{N}}$ as defined in
Eq.\ \ref{eq:lambda}. This gives us in Gaussian units

\begin{eqnarray}
\overline{V} & = & \alpha\,\sigma\,\phi\,I,\mbox{ with}\\
\alpha & = & \frac{3\,\alpha_{V}}{4\,\alpha_{\phi}\,\alpha_{I}\,B_{\bot0}^{1/2}}\nonumber \\
 & \approx & -4.269\cdot\sqrt{\frac{m_{\text{e}}^{3}\,c^{7}}{\text{e}^{5}\,\nu\,B_{0}}}\nonumber \\
 & \approx & \mbox{-2.189}\cdot10^{18}\left(\frac{\nu}{408\,\text{MHz}}\right)^{-\nicefrac{1}{2}}\left(\frac{B_{0}}{6\,\mu\text{G}}\right)^{-\nicefrac{1}{2}}
\end{eqnarray}
being a LOS-independent dimensionless quantity and
\begin{align}
\sigma= & \left(\int\!\!dl\,n_{\mathrm{th}}^{2}\overline{B}^{2}\right)^{-1}\times\nonumber \\
 & \left[\left(\int\!\!dl\,n_{\mathrm{rel}}\,\overline{B}^{2}\right)^{2}+2\lambda_{\text{N}}\int\!\!dl\,n_{\mathrm{rel}}^{2}\,\overline{B}^{4}\right]^{-1}\times\nonumber \\
 & \left[\frac{2}{3}B_{0}^{2}\left(\int\!\!dl\,n_{\mathrm{rel}}\,\overline{B}^{2}\right)\left(\int\,\,dl\,n_{\mathrm{th}}n_{\mathrm{rel}}\,\overline{B}^{2}\right)\right.\nonumber \\
 & \,\,\,\,\,\,\,+\,\left(\int\!\!dl\,n_{\mathrm{rel}}\,\overline{B}^{2}\right)\left(\int\!\!dl\,n_{\mathrm{th}}\,n_{\mathrm{rel}}\,\overline{B}^{4}\right)\nonumber \\
 & \left.\,\,\,\,\,\,\,+\,\,\,2\lambda_{\text{N}}\int\!\!dl\,n_{\mathrm{th}}\,n_{\mathrm{rel}}^{2}\,\overline{B}^{6}\right]
\end{align}
a LOS-dependent constant with dimension of an area. The unknown $\lambda_{\text{L}}$
canceled out and the unknown $\lambda_{\text{N}}$ affects only sub-dominant
terms, as it is e.g. compared in the denominator to the Galactic dimension
$L=\left(\int\!\!dl\,n_{\mathrm{rel}}\,\overline{B}^{2}\right)^{2}/\left(\int\!\!dl\,n_{\mathrm{rel}}^{2}\,\overline{B}^{4}\right)\gg\lambda_{\text{N}}$.
We therefore neglect terms proportional to $\lambda_{\text{N}}$ in
the following and calculate
\begin{eqnarray}
\sigma & \approx & \frac{\frac{2}{3}\,B_{0}^{2}\,\int\,\,dl\,n_{\mathrm{th}}n_{\mathrm{rel}}\,\overline{B}^{2}+\int\!\!dl\,n_{\mathrm{th}}\,n_{\mathrm{rel}}\,\overline{B}^{4}}{\left(\int\!\!dl\,n_{\mathrm{th}}^{2}\overline{B}^{2}\right)\,\left(\int\!\!dl\,n_{\mathrm{rel}}\,\overline{B}^{2}\right)}\nonumber \\
 & \approx & \frac{\frac{2}{3}\,\int\,\,dl\,x_{\mathrm{th}}^{1+\beta_{th}}x_{\mathrm{rel}}^{1+\beta_{rel}}+\int\!\!dl\,x_{\mathrm{th}}^{1+2\beta_{\mathrm{th}}}\,x_{\mathrm{rel}}^{1+2\beta_{rel}}}{n_{\mathrm{th}0}\left(\int\!\!dl\,x_{\mathrm{th}}^{2+\beta_{\mathrm{th}}}x_{\mathrm{rel}}^{\beta_{rel}}\right)\,\left(\int\!\!dl\,x_{\mathrm{th}}^{\beta_{\mathrm{th}}}x_{\mathrm{rel}}^{1+\beta_{rel}}\right)}\nonumber \\
\label{eq:sigma}
\end{eqnarray}
for each LOS to translate $d=\phi\,I$ into $\overline{V}$ there. 

\section{Prediction\label{sec:Prediction}}

\begin{figure*}
\includegraphics[bb=70bp 0bp 372bp 216bp,clip,width=0.5\textwidth]{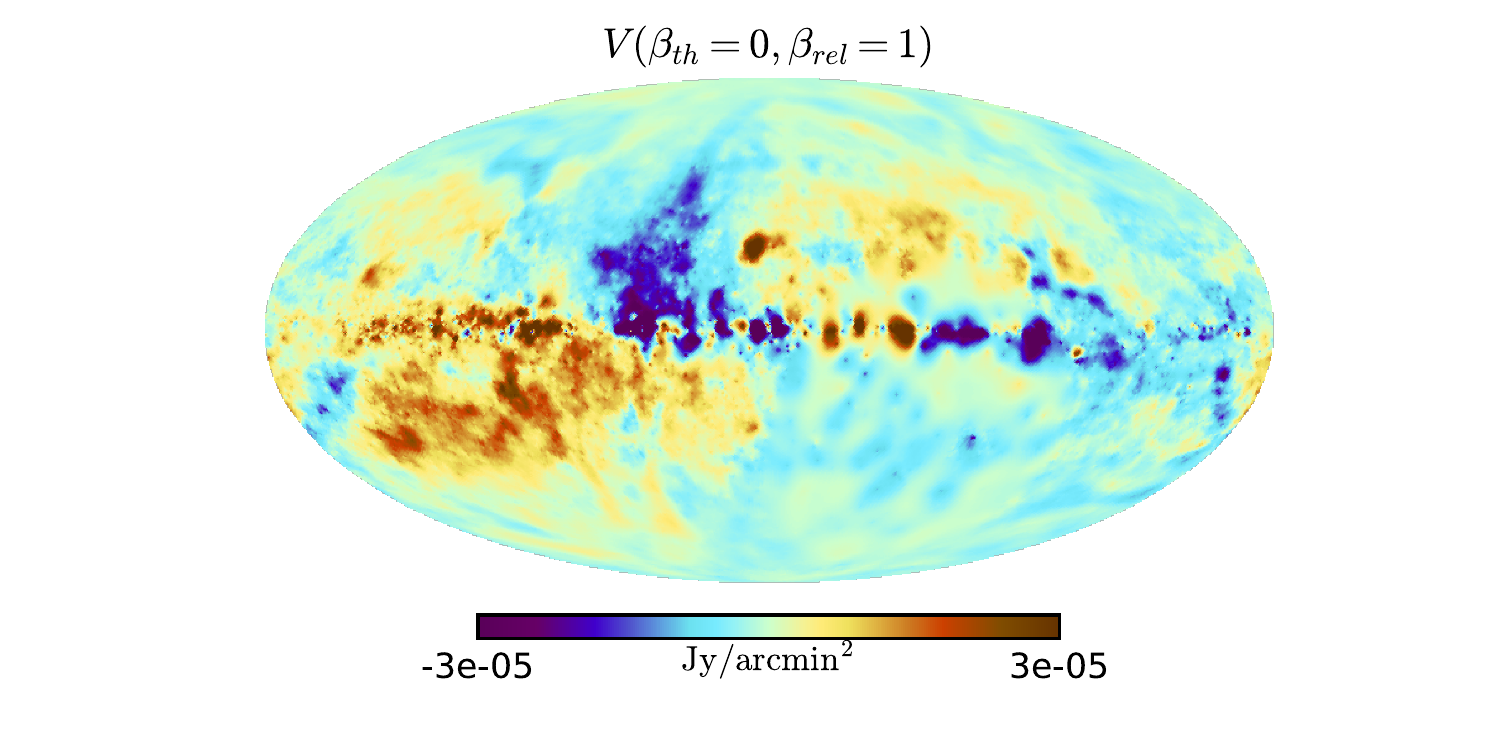}\includegraphics[bb=70bp 0bp 372bp 216bp,clip,width=0.5\textwidth]{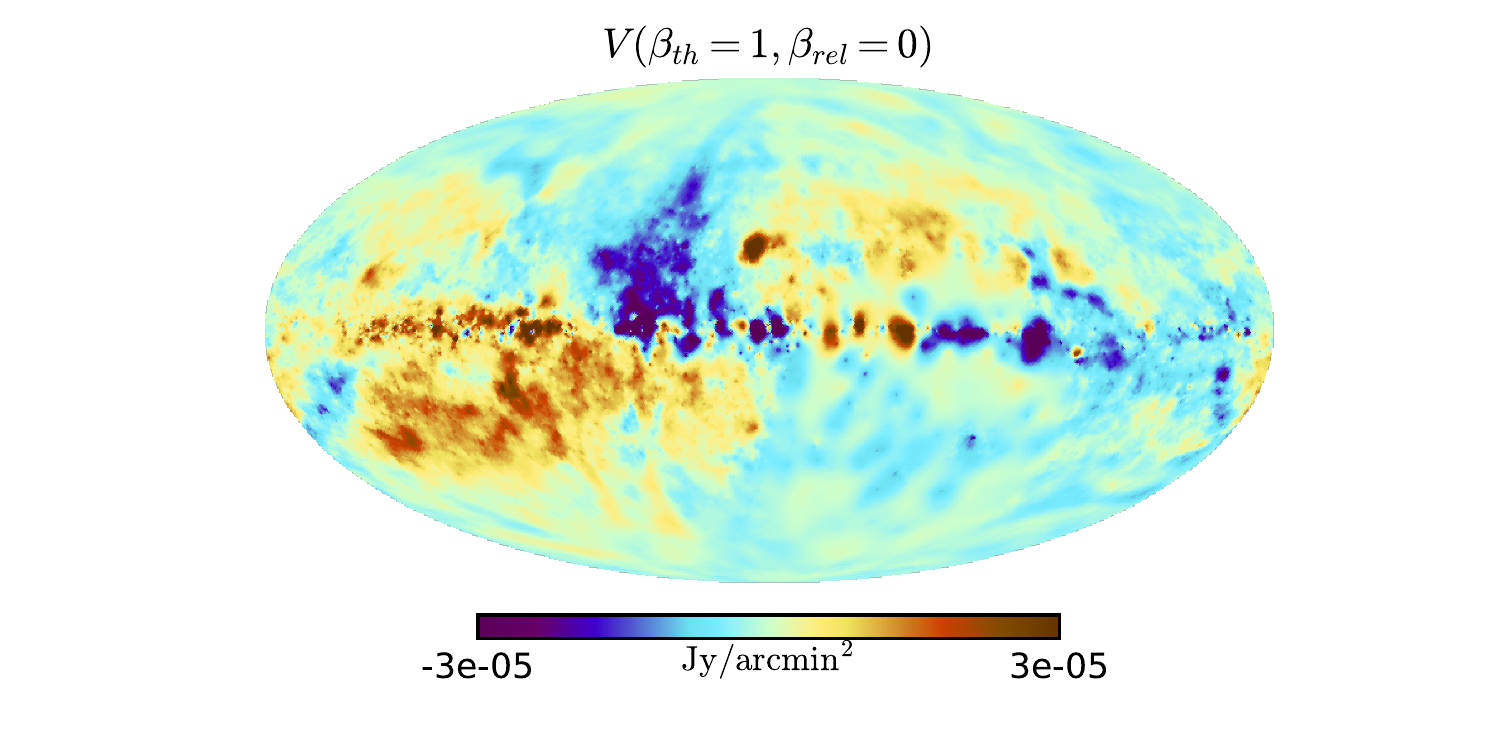}\caption{Predicted circular polarisation intensity at 408 MHz for $\beta=(0,1)$
(left) and $\beta=(1,0)$. Red indicates clockwise rotation, according
to the IAU convention for measuring angles that is opposite to the
mathematical convention.\label{fig:circ}}
\end{figure*}
\begin{figure*}
\includegraphics[bb=70bp 0bp 372bp 216bp,clip,width=0.5\textwidth]{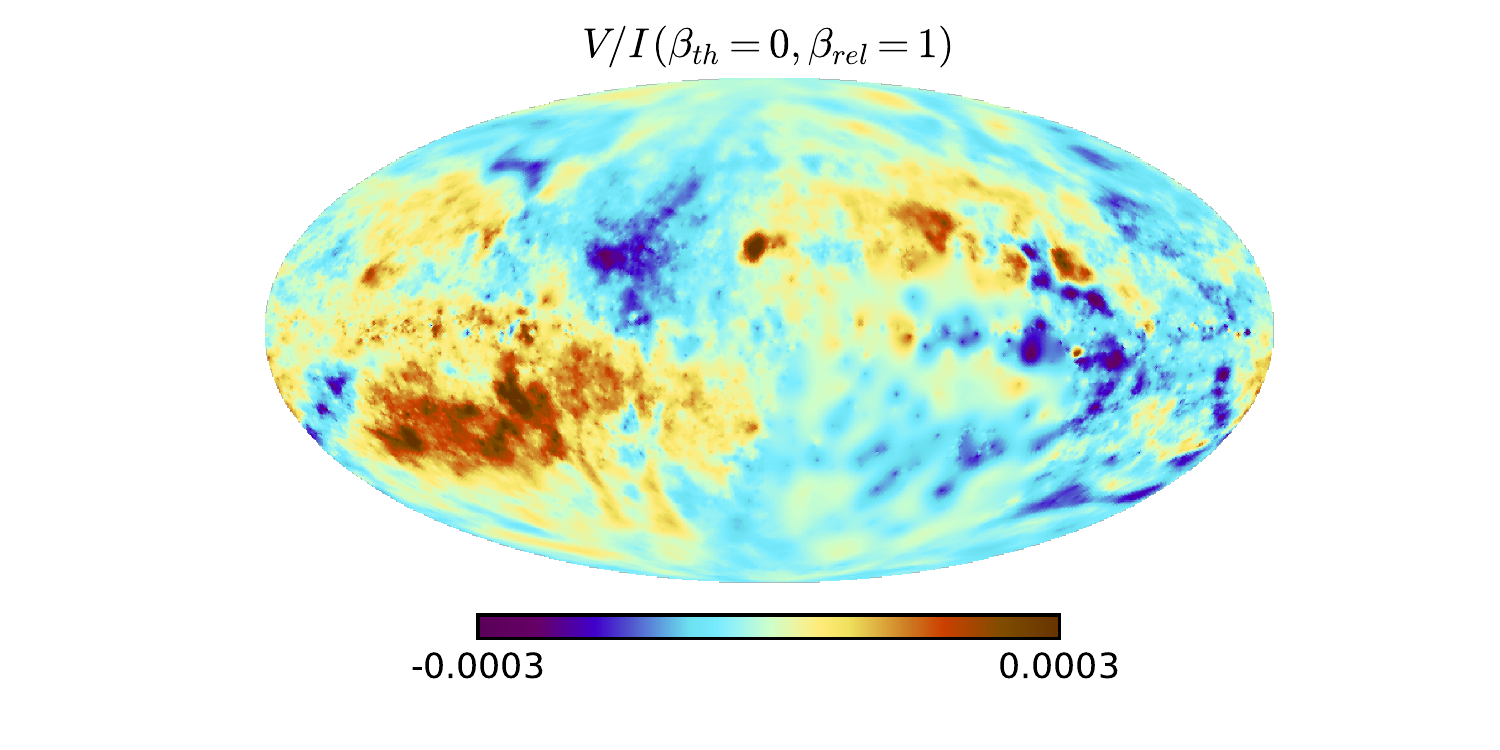}\includegraphics[bb=70bp 0bp 372bp 216bp,clip,width=0.5\textwidth]{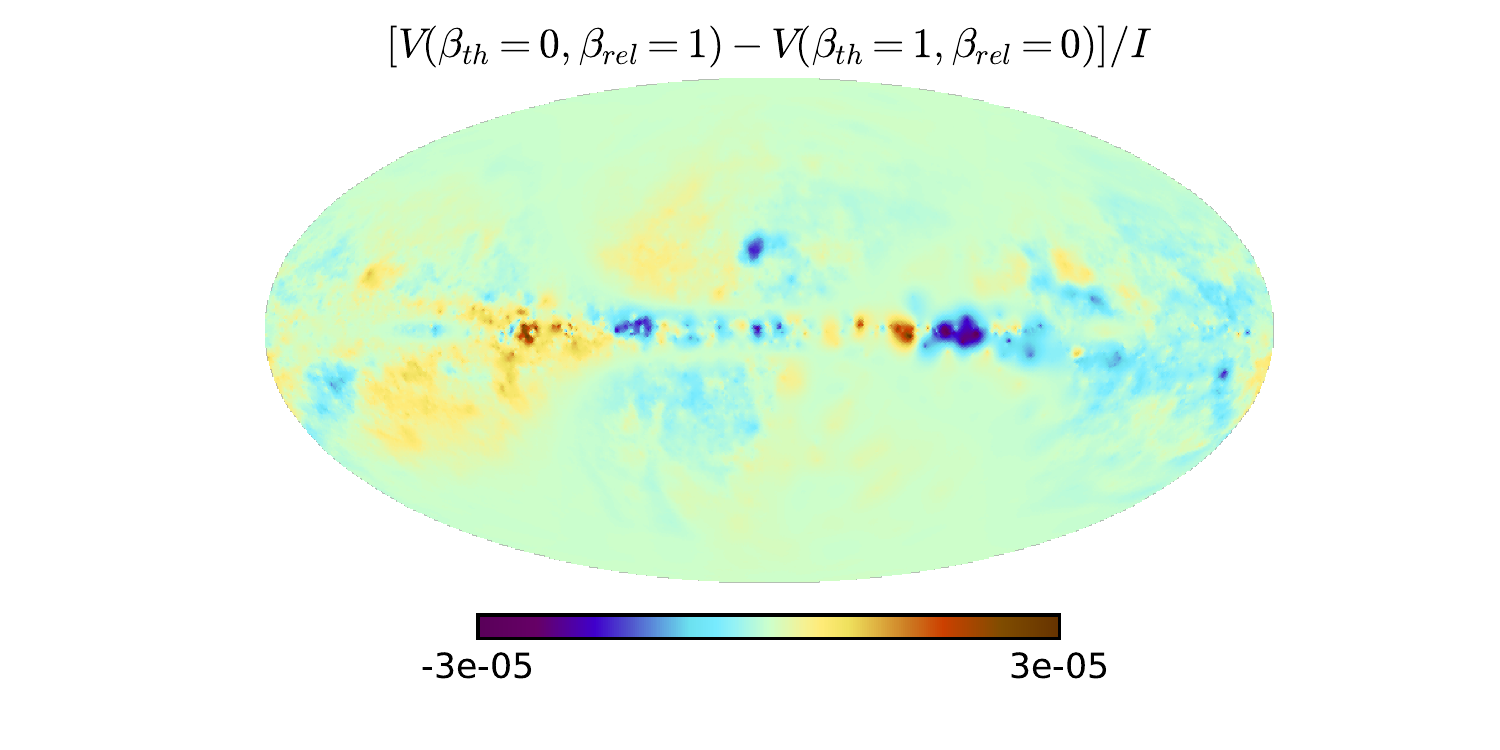}\caption{Predicted $V/I$ ratio at 408 MHz for $\beta=(0,1)$ (left) and the
difference of the $\beta=(0,1)$ and $\beta=(1,0)$ ratios (right)
.\label{fig:rel}}
\end{figure*}
To give an estimate for the CP sky, we need maps of the total synchrotron
intensity and the Faraday rotation of the Milky Way. We use the 408\ MHz
map provided by \citep{2015MNRAS.451.4311R}, which is based on the
data of \citep{1970MNRAS.147..405H,1974A&AS...13..359H,1981A&A...100..209H,1982A&AS...47....1H},
and the Faraday rotation map provided by \citep{2012A&A...542A..93O},
which is largely based on the data of \citep{2009ApJ...702.1230T}.
These are shown in Fig.\ \ref{fig:haslam}

We further need to quantify the $\sigma$ parameter given in Eq. \ref{eq:sigma}.
For this we need the thermal and relativistic electron distribution
of the galaxy and thereby $x_{\mathrm{rel}}$ and $x_{\mathrm{th}}$.
For the 3D distribution of the thermal electron density in the Milky
Way we use the NE2001 model \citep{2002astro.ph..7156C} without its
local features. The spatial and the energy distribution of relativistic
electrons in the Galaxy are more uncertain as we have only direct
measurements of the cosmic ray electrons near the Earth. Considerable
effort to infer these distributions have been made \citep{2001AdSpR..27..717S,2004ApJ...613..962S,2011CoPhC.182.1156V,2012A&A...542A..93O,2015arXiv150705958O,2016APh....81...21M,2017JCAP...02..015E}.
As we have shown in Eq. \ref{eq:sigma}, we only need the spatial
dependence and not the actual normalisation of $n_{\mathrm{rel}}$,
which means that this quantity only effects the relative strength
of different structures in the CP map and not the overall strength
of the predicted CP intensity itself. For this reason, and since we
only aim for a rough estimate, we are content with a simplistic large-scale
relativistic electron model. Given the distribution of matter in the
galaxy, a exponential model for the spatial structure of cosmic ray
electrons may make sense, as already adopted by other authors (\citep{2001ApJ...556..181D,2007ApJS..170..335P,2008A&A...477..573S,2010RAA....10.1287S}),
at least in a similar way. In our case, we can use Eqs. \ref{eq:I}
and \ref{eq:scaling-relation} to give an estimate of the of total
synchrotron map given our relativistic electron model and the scaling
parameters of Eq. \ref{eq:scaling-relation}, where we adopt $\beta=(0,1)$
and try to reproduce the large scale pattern of the 408 MHz map shown
in Fig. \ref{fig:haslam}. We thereby choose the following model for
the spatial dependence of the relativistic electrons: 

\begin{equation}
x_{\mathrm{rel}}=e^{-\vec{|r|}/r_{0}}\cdot\mathrm{cosh^{-2}}(|\vec{z|}/z_{0})\label{eq:x_rel}
\end{equation}
The vector $\vec{r}$ points in the radial direction in the galactic
plane, the vector $\vec{z}$ points out of the plane. As mentioned
before, the parameters $r_{0}$ and $z_{0}$ are estimated via a naive
comparison of the observed and estimated synchrotron maps at 408 MHz
shown in Figs. \ref{fig:haslam} and \ref{fig:i_predict_01}, respectively.
The parameters adapted in this work are $r_{0}=12\,\mathrm{kpc}$
and $z_{0}=1.5\,\mathrm{kpc}$. Given the morphological complexity
of the map in relation to the simplicity of the model and the poorly
understood nature of the origin and evolution of electron cosmic rays
we acknowledge that the parameters of this model are highly uncertain.
Also completely different parametrization of $x_{\mathrm{rel}}$ might
lead to the same estimate for $I$ because of the projection involved.
The conversion factor $\alpha\,\sigma$ implied by our rough 3D model
at 408 MHz is also shown in \ref{fig:i_predict_01} for $\beta=(0,1)$.

\begin{figure*}
\includegraphics[bb=0bp 0bp 600bp 400bp,clip,width=0.5\textwidth]{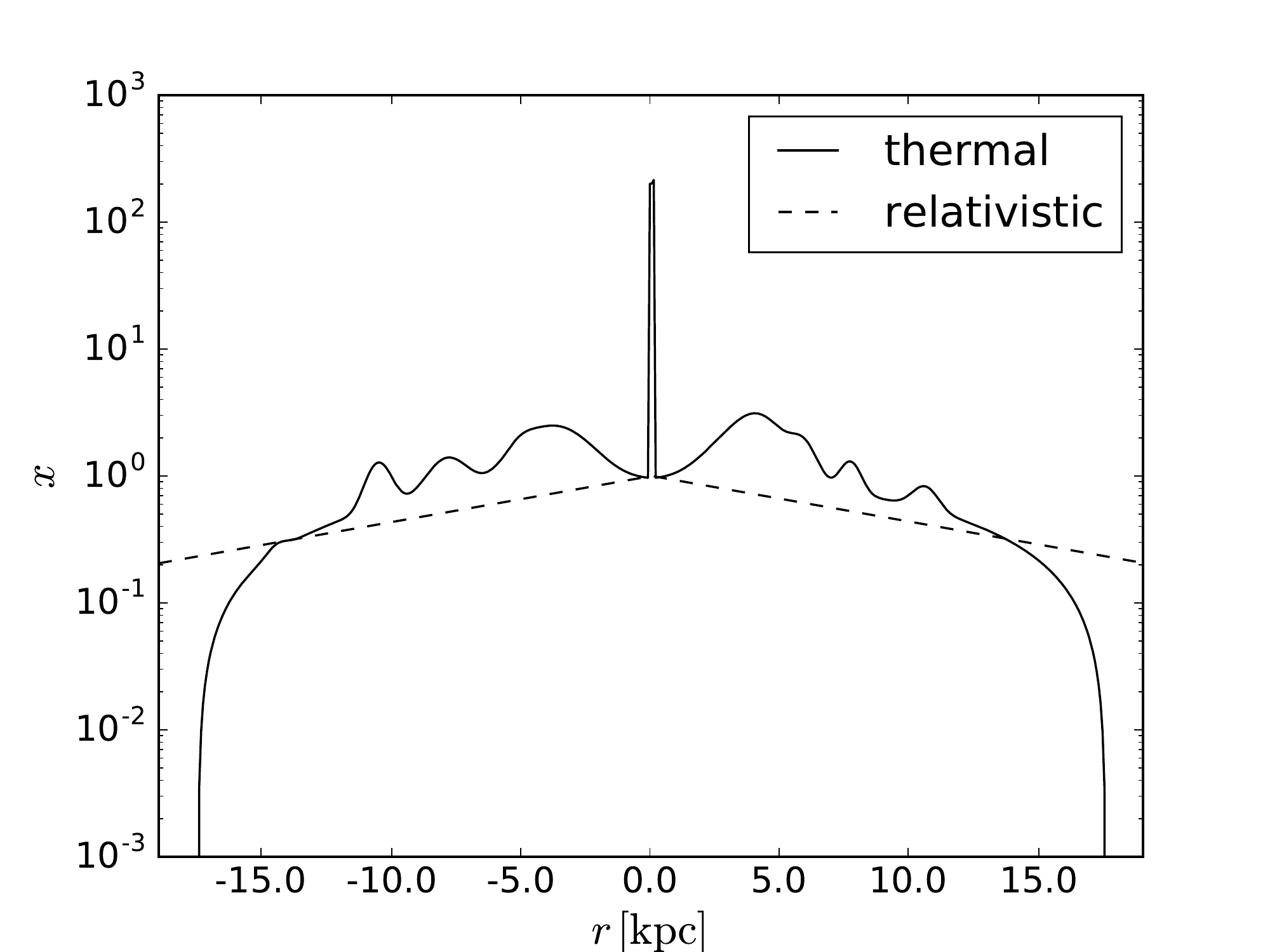}\includegraphics[bb=0bp 0bp 600bp 400bp,clip,width=0.5\textwidth]{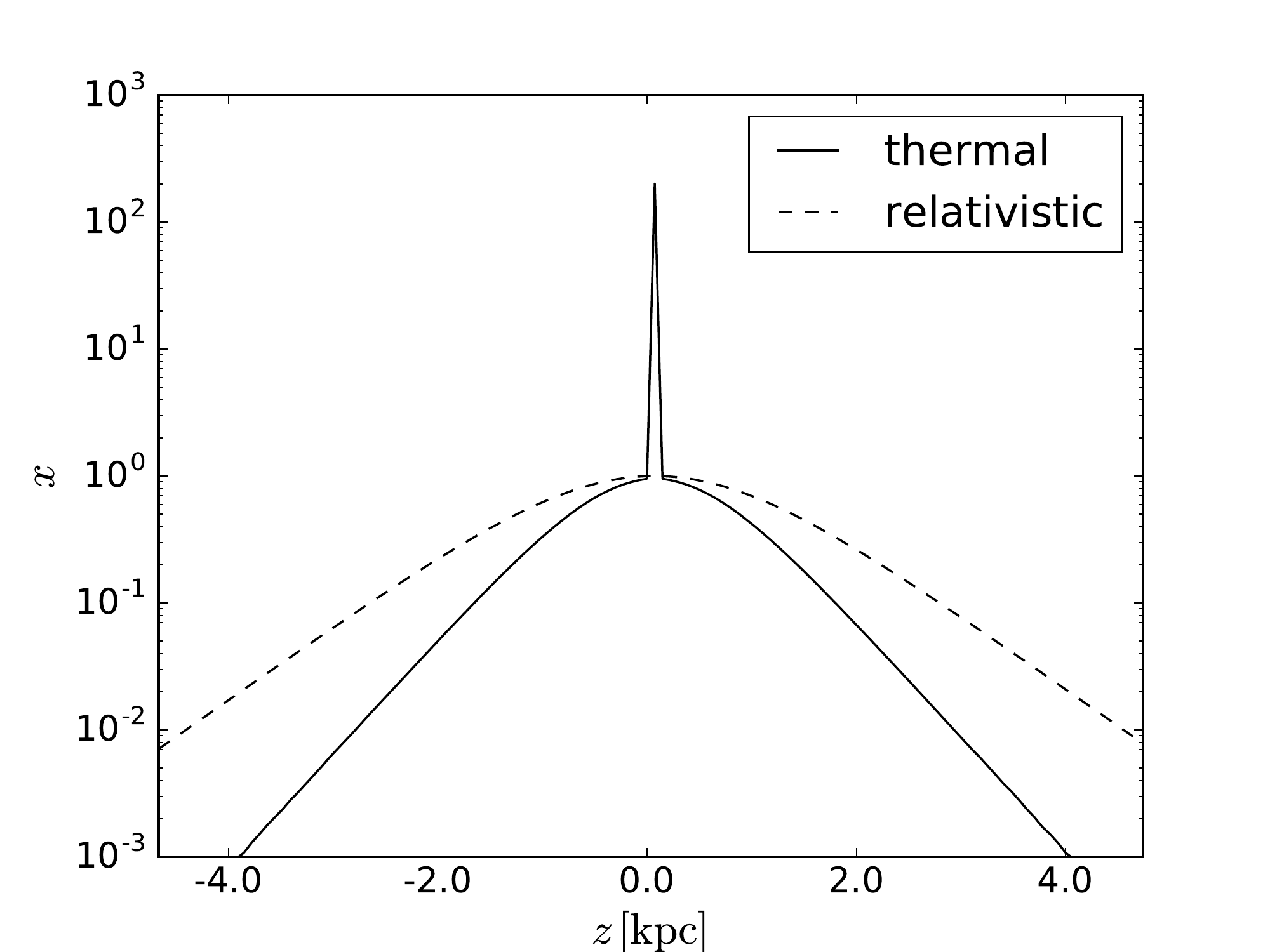}\caption{Profiles of the thermal and relativistic electron density used in
this work in terms of the dimensionless quantities $x_{\mathrm{\mathrm{th}}}$
and $x_{\mathrm{rel}}$ as defined in the context of Eq. \ref{eq:scaling-relation}.
\label{fig:el_densities}}
\end{figure*}

The resulting estimate of the circular polarisation intensity of the
Milky Way is depicted in Figs. \ref{fig:circ} for the two cases $\beta=(0,1)$
and $\beta=(1,0)$. The morphology of the resulting maps is dominated
by the morphology of the Faraday and the synchrotron map, what seems
natural given our formalism. The influence of the dependence of the
magnetic field on the different electron densities seems to be small,
as the difference between between the two complementary cases is negligible,
as we show for the predicted $V/I$ ratio in Fig.\ \ref{fig:rel}.
We predict a signal of up to $5\cdot10^{-4}$ Jansky per square arcminute
at $408\,\mathrm{MHz}$ and more at lower frequencies. The CP is strongest
in the center plane of the Galaxy. The relative strength of the CP
intensity to the total synchrotron intensity up to $V/I\sim3\cdot10^{-4}$
as depicted in Fig. \ref{fig:rel}. The $V/I$ ratio is largest just
above and below the disc, as well as in some spots in the outer disc.
We expect this ratio to increase with $\nu^{-0.5}$, approaching $10^{-3}$
at $40$ MHz, which might be a detectable level for current instrumentation\citep{2017arXiv170604200M}.
The frequency scaling of $V/I\propto\nu^{-0.5}$ was already predicted
by \citep{2016PhRvD..94b3501K} for the GHz range.

The diffusion length of relativistic electrons depends on energy,
therefore, the radio sky at different frequencies is not just a rescaled
version of the 408\ MHz map used as a template here. The $V/I$ map
provided by this work, however, should \textendash{} within its own
limitations \textendash{} be valid at others frequencies as well.
Therefore, it can be used after scaling by $(\nu/\text{408 MHz})^{-0.5}$
to translate total intensity templates at other frequencies into CP
expectation maps at the same frequency, which then incorporate any
difference of the radio sky due to spatially varying relativistic
electron spectra.

Anyhow, even if a total intensity template is not available at the
measurement frequency, the main structure of the CP prediction, which
are the sign changes induced by the sign changes of the Faraday sky,
will be robust with respect to a change in frequency. Therefore, the
CP template should be used as a structure expected on the sky, while
allowing the real sky to deviate by some factor from it due to errors
induced by the assumed frequency scaling and other simplifications.
A template search method that is robust in this respect, is discussed
below.

The assumed scaling of the magnetic field energy density with the
electron densities, $\beta$ has only a minor impact on the result.
The difference between the $\beta=(0,1)$ and the $\beta=(1,0)$ scenarios
is less than 10\%, as Fig.\ \ref{fig:rel} shows. Together with Fig.
\ref{fig:el_densities} this is indeed evidence for the robustness
of our results, as the profiles of relativistic and thermal electrons
used in this work are quite different, nonetheless the different scaling
does not lead to significantly different CP maps. 

\section{Detection strategy\label{sec:Detection-strategy}}

\subsection{Traditional imaging}

Now, we investigate the possibility to detect this CP signal with
single-dish and interferometric observations, by requiring a signal-to-noise
ratio of 10 in CP and by assuming a Stokes-$V$ to $I$ sensitivity
ratio of $\sigma_{V}/\sigma_{I}\approx\sqrt{2}$ and a power-law total
intensity frequency spectrum ($I_{\nu}\propto\nu^{\alpha}$, with
$\alpha=-0.8$). For example, the Sardinia Radio Telescope has the
capability to observe in the low portion of the electro-magnetic spectrum:
in P-band ($305-410\,$\ MHz) and in L-band ($1.3-1.8\,$\ GHz).
By using the specifications given in \textbf{\citep{2017arXiv170309673P}},
an observing time of $\lesssim1\,$s per beam is required to reach
the requested sensitivity in both frequency bands. 

One of the largest surveys of the sky at the moment available is the
NRAO VLA Sky Survey (NVSS, \textbf{\citep{1998AJ....115.1693C}})
characterized by a sensitivity in Stokes $U$ and $Q$ of $0.29$\ mJy/beam.
The sensitivity at this frequency and resolution to detect the signal
we are interested in is $\approx6\,\mu$Jy/beam. In principle, such
a sensitivity can be reached by stacking all the 2326 fields of the
survey if one can assume $\sigma_{{\rm V}}\approx\sigma_{{\rm Q}}\approx\sigma_{{\rm U}}$.

We performed a similar evaluation for the LOFAR and the SKA, by referring
to the lowest frequency band available for these instruments. For
LOFAR, we use $45$\ MHz, where we expect $V/I\approx0.001$. At
this frequency, the required sensitivity is reached in less than 1\ s.
For the SKA, we considered the SKA-Low specifications given after
the re-baselining in the frequency range $50-350$\ MHz, with a central
frequency of $200$\ MHz and a bandwidth of $300$\ MHz. The required
sensitivity can be reached in 3\ h of observing time, if a resolution
of $\approx7$ arcsec is considered. The Effelsberg telescope should
obtain enough sensitivity within 20 minutes observation in its 400\ MHz
band and the GMRT within 30\ min in its 200\ MHz band. 

Thus, the prospects to detect the predicted CP signal are good from
a pure signal to noise perspective. However, the polarization accuracy
after calibration of the new generation of radio telescopes is typically
of $0.1-1$\ \% \citep[e.g.][]{2013ApJS..206...16P,2016MNRAS.461.3516M}.
This instrumental limitation will make the imaging of the CP signal
extremely hard as contamination of the CP signal by polarization leakage
will be in the best case as strong as the signal we predict, in many
cases one or two orders of magnitude stronger. 

To overcome this, we propose to cross-correlate the measured CP sky
with our predicted one, as such instrumental effects are not present
in our prediction and therefore should statistically averaged out
in the comparison. 

\subsection{Template search}

The CP all sky prediction constructed in the previous section can
be used to search for the weak Galactic CP signal even in strongly
contaminated data. Although CP sky images are usually not available,
a number of radio telescopes take circular polarization data 
\begin{align}
d_{V} & =\int_{\mathcal{S}^{2}}d\hat{n\,}R(\hat{n})\,V(\hat{n})+\xi.
\end{align}
Here, $d_{V}=(d_{V1},\,\ldots d_{Vu})\in\mathbb{C}^{u}$ is the data
vector of length $u$. $\hat{n}$ is a direction on the celestial
sphere $\mathcal{S}^{2}$. $R:\mathcal{S}^{2}\rightarrow\mathbb{C}^{u}$
is the CP instrument response encoding the primary beam, the Fourier
transform of the sky and subsequent sampling in case of interferometers,
and any gain factors of the telescope. $V(\hat{n})$ is the CP sky
and $\xi\in\mathbb{C}^{u}$ is the noise vector of the observation
including the cross talk from other Stokes parameters. Here, we assume
$\xi$ to be generated by a zero-mean stochastic process with known
covariance $\Xi=\langle\xi\,\xi^{\dagger}\rangle_{(\xi)}$, which
has to be obtained by careful studying the instrumental properties.

Some part of the observed data vector can now be predicted using the
CP prediction$\overline{V}(\hat{n})$, namely 
\begin{equation}
\overline{d_{V}}=R\,\overline{V}=\alpha\,\int_{\mathcal{S}^{2}}d\hat{n\,}R(\hat{n})\,\sigma(\hat{n})\,\phi(\hat{n})\,I(\hat{n}).
\end{equation}
Since our prediction might be off by some multiplicative factor due
to the various approximations involved in its derivation, and since
we did not attempt to calculate the model uncertainty, a comparison
via a likelihood function $\mathcal{P}(d_{V}|\overline{d_{V}})$ is
out of reach. However, a simple, but sensitive indicator function
(or test statistics) for the presence of the predicted CP signal is
the inversely noise-weighted scalar-product of observed and predicted
data,
\begin{equation}
t=\overline{d_{V}}^{\dagger}\Xi^{-1}\,d_{V}.
\end{equation}
If $V=\gamma\,\overline{V}+\delta V$ is the correct CP sky, with
$\gamma\sim1$ the factor necessary to correct for our approximations,
$\delta V$ the CP structures missed by our prediction due to imperfect
correlation of $V$ with $d=\phi\,I$, and $U=\langle\delta V\,\delta V^{\dagger}\rangle_{(\vec{B},n)}$
the imperfection covariance , we expect
\begin{eqnarray}
\overline{t} & = & \langle t\rangle_{(\vec{B},\xi|n)}=\gamma\,\overline{V}^{\dagger}R^{\dagger}\Xi^{-1}R\,\overline{V}>0\mbox{ and}\nonumber \\
\sigma_{t}^{2} & = & \langle(t-\overline{t})^{2}\rangle_{(\vec{B},\xi|n)}=u+\mathrm{Tr}\left[U\,R^{\dagger}\Xi^{-1}R\right]
\end{eqnarray}
and therefore a signal-to-noise ratio (SNR) of 
\begin{align}
\frac{S}{N}=\frac{\overline{t}^{2}}{\sigma_{t}^{2}} & =\frac{\gamma^{2}\,\left(\mathrm{Tr}\left[R\,\overline{V}\,\overline{V}^{\dagger}R^{\dagger}\Xi^{-1}\right]\right)^{2}}{u+\mathrm{Tr}\left[R\,U\,R^{\dagger}\Xi^{-1}\right]},
\end{align}
where we used $\mathrm{Tr}\left[\Xi\,\Xi^{-1}\right]=u,$ the number
of data points. If we only reconstructed $f=10\%$ of the intensity
of the true celestial CP signal, so that $\gamma^{2}\langle\overline{V}\,\overline{V}^{\dagger}\rangle\approx f^{2}\,U=10^{-2}\,U$,
and if the CP data is $99\%\,(=1-p)$ noise and cross leakage dominated,
so that $R\,\langle V\,V^{\dagger}\rangle R^{\dagger}\approx R\,U\,R^{\dagger}\approx p^{2}\,\Xi\approx10^{-4}\,\Xi$,
we get a SNR of $\nicefrac{S}{N}\approx f^{2}\,p^{2}\,u=10^{-12}\,u$
and enter the detection range ($\nicefrac{S}{N}\sim1$) in the terabyte
regime ($u\sim f^{-2}\,p^{-2}=10^{12}$). 

\section{Conclusions\label{sec:Conclusions}}

Using the observational information on magnetic fields along and perpendicular
to the LOS from Faraday rotation and synchrotron total emission we
provided a detailed map of the expected diffuse Galactic CP emission\footnote{The $V$ and $V/I$ maps for the two scenarios discussed are available
at \url{http://wwwmpa.mpa-garching.mpg.de/ift/data/CPol/}.}, which is at a level of $3\cdot10^{-3}$ of the total intensity at
408\ MHz and higher at lower frequencies. This prediction relies
on assumptions about the magnetic field statistics, and the three
dimensional distributions of thermal and relativistic electrons throughout
the Milky Way. As these assumptions are not certain, the real Galactic
CP sky can and will differ from our prediction. Nevertheless, the
provided CP prediction can be used for template based searches for
the elusive CP signal. Our model shows similarities and differences
to a CP prediction based on a 3D models of the Milky Way \citep{2016PhRvD..94b3501K}.
We expect our model to capture more details of the real CP sky, as
its construction is based directly on observed data sets, without
the detour of using those to construct a parametrized, and therefore
coarse, 3D model. However, which model is more accurate should certainly
be answered by observations.

A confirmation of the celestial CP signal we predict would indicate
a co-location of the origin of the observed Faraday rotation signal
and synchrotron emission. In case the predicted signal is not detectable
with a strength comparable to the prediction, this would indicate
a spatial separation of these regions along the LOSs and therefore
important information on the Galactic magnetic field structure and
its correlation with thermal and relativistic electrons. 

Finally, we like to point out that the hypothetical possibility exist
that the observed CP signal has the opposing sign compared to our
prediction (even after potential confusions of the used CP conventions
are eliminated). This would happen in case the synchrotron emission
of the Milky Way would predominantly result from relativistic positrons,
which gyrate in the opposite direction compared to the electrons.
This is \textendash{} however \textendash{} very unlikely given that
the observed local density of relativistic electrons is much higher
than that of the positrons and given that the relativistic particles
in the Milky Way are believed to be accelerated out of the thermal
particle pool. Nevertheless, it shows that the charge of the Galactic
synchrotron emitters can actually be tested by sensitive CP observations. 
\begin{acknowledgments}
This research was supported by the DFG Forschengruppe 1254 \textquotedblleft Magnetisation
of Interstellar and Intergalactic Media: The Prospects of Low-Frequency
Radio Observations\textquotedblright . We thank Henrik Junklewitz
and Rick Perley for discussions and an anonymous referee for constructive
feedback. 
\end{acknowledgments}

\appendix
\bibliographystyle{apsrev}
\bibliography{CPol}

\end{document}